# Modeling and simulation of non-linear and hysteresis behavior of magneto-rheological dampers in the example of quarter-car model

Sulaymon Eshkabilov
Dynamics & Control Lab
Tashkent Automotive Road Institute
Amir Temur Str. 20, Tashkent -100060, Uzbekistan
Email: sulaymon@d-c-lab.com

***Abstract.*** *This paper presents reviews of several models and numerical simulation models of non-linear and hysteresis behaviors of magneto-rheological liquid dampers in MATLAB®/Simulink® in the example of quarter-car model of vehicle suspension simulation, such as, Bingham, Dahl, LuGre and Bouc-Wen models. In addition, it demonstrates numerical simulation models built in MATLAB®/Simulink® and discusses results from numerical simulation models for three different input excitations from terrain.*

1. **Introduction**

In general, most of the natural phenomena, operational machine processes and dynamic system behaviors are of non-linear nature that is very often linearized for the sake of simplicity in formulations and analyses. In fact, nonlinear behaviors or phenomena of processes may create difficulties in studies and engineering design processes but considering some of those non-linear characteristics of processes or behaviors of dynamic systems carefully could be also very beneficial and of great importance for efficient and accurate control, and used for operational efficiency and energy preservation or dissipation depending on their application areas. For example, nonlinear parameters and characteristics of some materials and interactions of different parts made of different materials have a great potential to apply for dampers and shock absorbers [1]. One of the good examples for such processes is a hysteresis loop observed in magnetic or magnetized materials and magneto-rheological (MR) liquids. In studies [2, 3, 4, 5], the MR liquids are found to be one of the most suitable and promising in designing vibration dampers and shock absorbers, and there are some combinatorial designs [6] of MR fluid dampers. In studies [7], feasibility of MR liquid damper modeling by employing Bouc-Wen model in association with an intelligent self-tuning PID controller for semi-active suspension modeling is studied numerically via computer modeling in MATLAB/Simulink. Nevertheless, identification of the hysteresis loop parameters is rather complex and may require considerable laboratory and numerical studies in order to apply them and get a best use of MR damper properties.

In this paper, we put some emphases on different mathematical models and formulations of the MR liquids, and their hysteresis loop parameters and numerical simulation models designed for a semi-actively controlled feedback damper for a vehicle suspension systems developed in MATLAB/Simulink. In addition, we shall try to analyze and compare efficiency and accuracy of these models in the example of the quarter-car model to design a semi-active suspension system.

2. **Mathematical formulation of a quarter-car model**

To derive an equation of (vertical) motion of a vehicle while driving on uneven roads, we take quarter of a vehicle by assuming that terrain roughness is evenly distributed under all wheels of a vehicle and loading from the whole vehicle body is equally distributed across all of its axles. In addition, we consider that a tire has some damping effect. With these preconditions, we draw the next physical model (Figure 1) of the system for passively and semi-actively controlled systems of a quarter-car model.

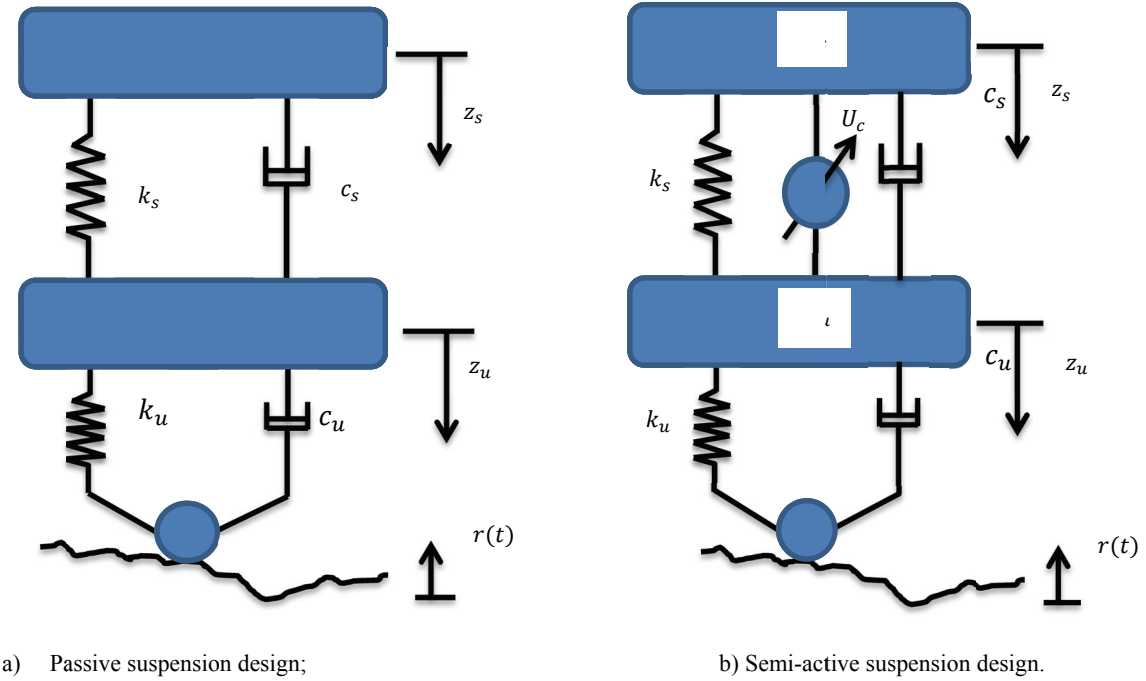

a) Passive suspension design;  b) Semi-active suspension design.

Figure 1. Vehicle suspension models.

From the passive and semi-active suspension design shown in Figure 1, we can derive equations of motion of the two mass bodies which as un-sprung mass (half of axle mass and one wheel) $m_u$ and sprung mass (quarter car body mass) $m_s$. So, the equations of motion of the systems are

a) For passive suspension system:
$$\begin{cases} m_s\ddot{z}_s + c_s(\dot{z}_s - \dot{z}_u) + k_s(z_s - z_u) = 0 \\ m_u\ddot{z}_u + c_s(\dot{z}_u - \dot{z}_s) + k_s(z_u - z_s) + c_u\dot{z}_u + k_u z_u = k_u r + c_u \dot{r} \end{cases} \quad (1)$$

b) For semi-active suspension system:
$$\begin{cases} m_s\ddot{z}_s + c_s(\dot{z}_s - \dot{z}_u) + k_s(z_s - z_u) = U_c \\ m_u\ddot{z}_u + c_s(\dot{z}_u - \dot{z}_s) + k_s(z_u - z_s) + c_u\dot{z}_u + k_u z_u = -U_c + k_u r + c_u \dot{r} \end{cases} \quad (2)$$

Where $z_s, \dot{z}_s$ and $\ddot{z}_s$ are displacement, velocity and acceleration of the sprung mass (quarter car body mass), respectively; $z_u, \dot{z}_u$ and $\ddot{z}_u$ are displacement, velocity and acceleration of the un-sprung mass (half of axle mass and one wheel), respectively; $c_s$ and $c_u$ damping coefficients of suspension and tire; $k_s$ and $k_u$ stiffness of suspension and tire; $r(t)$ and $\dot{r}$ are terrain roughness (disturbance) displacement and velocity with respect to longitudinal speed of the vehicle; $U_c$ is the force generated by the controller that takes into account terrain roughness $r(t)$, and vertical displacement and velocity of the vehicle. In the model, for $U_c$ - the control force exerted by the controller, we apply several different hysteresis effect models, such as, Bingham, Dahl, LuGre and Bouc-Wen models and design numerical simulation models in MATLAB/Simulink.

### 3. Mathematical formulations of the MR dampers
### 3.1. Bingham model

To simulate and identify parameters of the MR liquids, Bingham plastic model [8] was proposed in 1985. It is formulated by the following:

$$F_{mr} = F_c \, sgn(\dot{y}) + c_0 \dot{y} + F_0 \quad (3)$$

Where $y$ is a piston's relative displacement and $\dot{y}$ is its derivative that is velocity of a piston; $F_c$ is frictional force; $c_0$ is damping constant; $F_0$ is offset force (constant force value). The signum function $sgn(\dot{y})$ will take care of the direction of the frictional force $F_c$ depending on the relative velocity $\dot{y}$ of the hysteresis (internal) variable $y$. Note that in our simulation model, $y$ and $\dot{y}$ correspond to the displacement $z$ and velocity $\dot{z}$ of the sprung mass.

The response of Bingham model corresponds to the following graph shown in Figure 3 and it can be assumed that the shape of Bingham model force $F_{mr}$ will be equal to Coulomb force plus friction force ($F_c$). The damping coefficient (constant) $c_0$ will be equal to the linear relationship between the force $\Delta F$ and the velocity $\Delta \dot{z}$ differences- Figure 3.

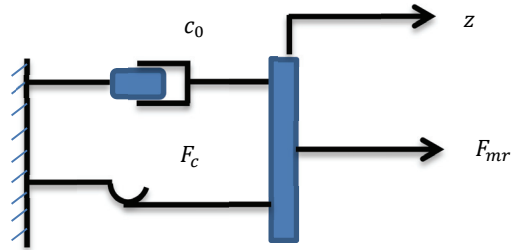

Figure 2. Bingham mechanical model proposed by [6].

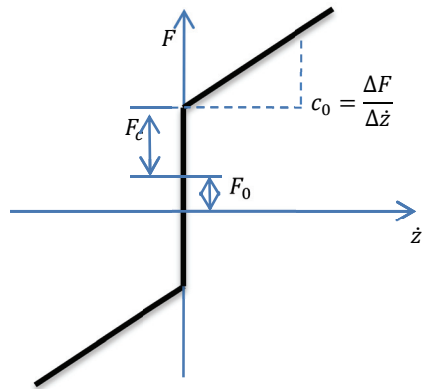

Figure 3. The response of Bingham model.

Now we build a Simulink model – Figure 4 using the formulation from the equation (3) and link it with the model expressed for the semi-actively controlled suspension model from the expression of (2) as shown in Figure 1.b.

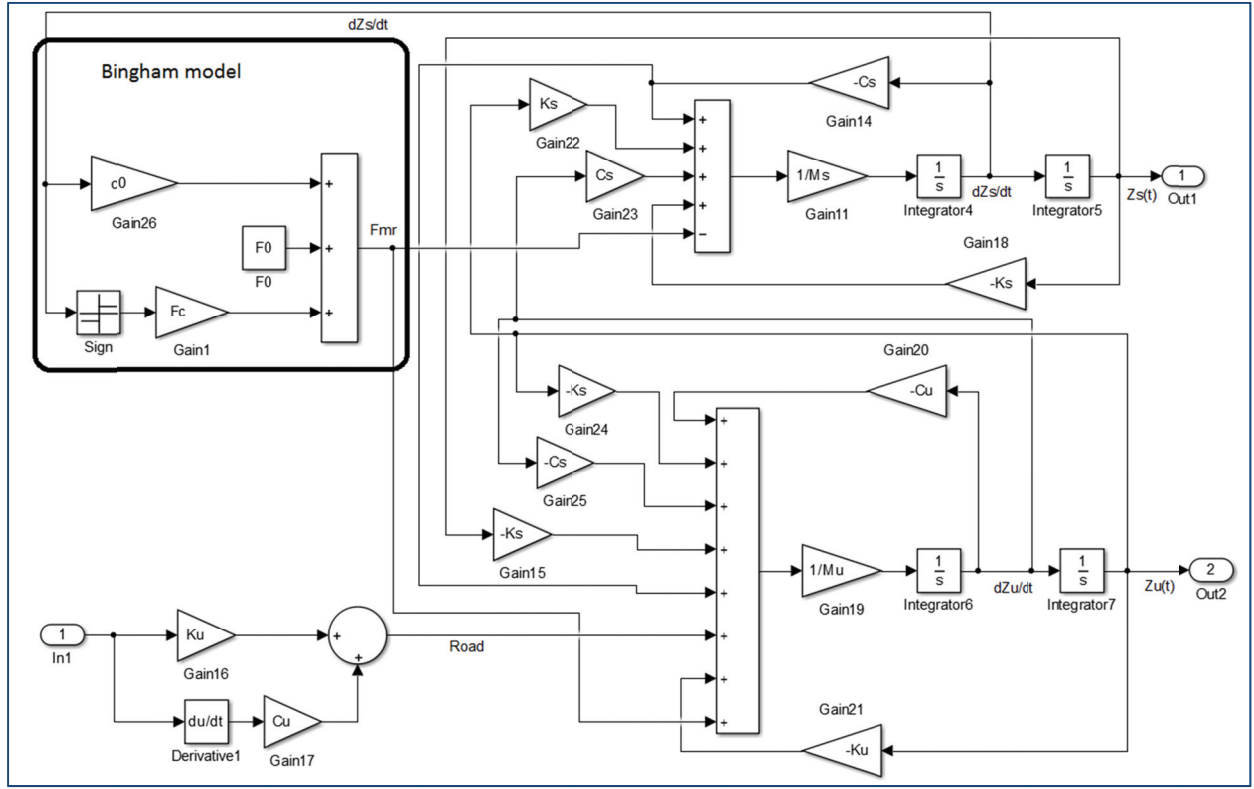

Figure 4. Bingham model embedded in semi-active suspension control.

Note that the Coulomb frictional force ($F_c$) is directly related to the yield stress. In Bingham model there are two input signals, which are $\dot{z}_u$ and $\dot{z}_s$ make up relative velocity $\dot{z} = \dot{z}_s - \dot{z}_u$ in order to direct the Coulomb frictional force $F_c$ with the signum function, i.e., $sgn(\dot{z})$.

**3.2. Dahl model**

This model considers quasi-static bonds in the origin of friction [9]. Dahl model of the MR damper [8] is formulated by:

$$F_{mr} = k\dot{z} + (k_{wa} + k_{wb}v)w \qquad (4)$$

$$\dot{w} = \rho(\dot{z} - |\dot{z}|w) \qquad (5)$$

Where, $F_{mr}$ is exerted force from the MR damper, $v$ is the control voltage, $w$ is the dynamic hysteresis coefficient, $k, k_{wa}, k_{wb}$ and $\rho$ are parameters that control the hysteresis loop shape.

Using the expressions (4) and (5), we build a simulation model of Dahl model in Simulink as shown in Figure 5. In Dahl model (Figure 5), there is one feedback coming from the un-spring mass that is velocity $dz(t)$ and there is one output signal that is $F_{mr}$ going to the un-sprung mass and sprung mass.

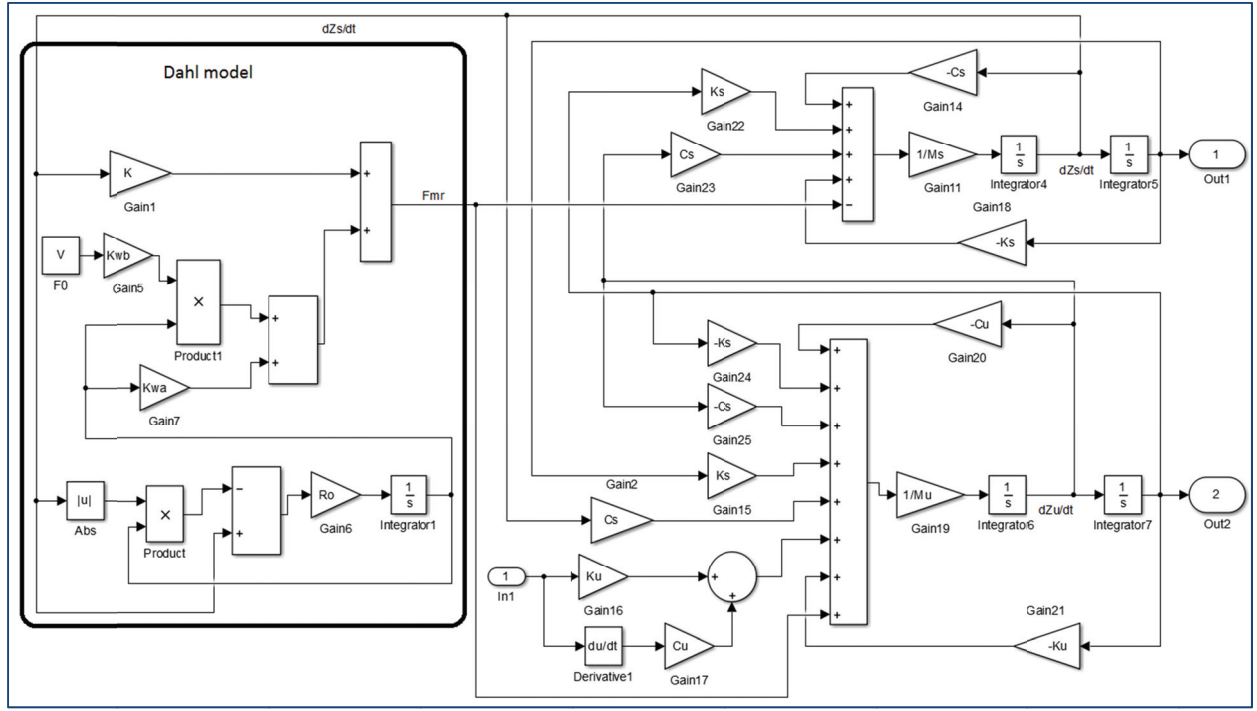

Figure 5. Dahl model implemented for semi-active control of suspension system.

In Dahl model, there are one input signal $\dot{z}_s$ and output signal $F_{mr}$. The input signal is coming from the velocity of the sprung mass that is a car body velocity and feeding summing junction of $F_{mr}$ force, and summing junction of $w$ dynamic hysteresis coefficient. The output signal is the control force feeding a summing junction of input forces for sprung mass with (-) sign and for un-sprung mass with (+) sign.

### 3.3. LuGre model

In modeling the hysteresis loops, the LuGre model is developed within studies [10] and applied in works [11] in modeling and simulation of dampers. This model [10] takes into account three types of frictions observed in dry friction and fluid flows, viz. Coulomb, stick-slip and stribeck effects that are formulated by the following:

$$F_{mr}(t) = \sigma_0 y(t) + \sigma_1 \dot{y}(t) + \sigma_2 \dot{z}(t) \tag{6}$$

Where $\sigma_0, \sigma_1, \sigma_2$ are stiffness, damping and viscous friction coefficients, respectively; $y(t)$ is the friction state (average deflection of the bristles), $\dot{y}(t)$ is the velocity of the friction state, $\dot{z}(t)$ is the relative velocity of the sprung mass.

$$\dot{y}(t) = \dot{z}(t) - \frac{|\dot{z}(t)|}{y_{ss}(\dot{z}(t))} y(t) \tag{7}$$

In the above expression, $y_{ss}(\dot{z}(t))$ is defined by [12 and 13] that has been expressed with the following

$$y_{ss}(\dot{z}(t)) = \frac{1}{\sigma_0}\left(F_c + (F_s - F_c)e^{-\left(\frac{\dot{z}(t)}{v_s}\right)^2}\right) \tag{8}$$

Where $F_c$ is the Coulomb friction force, $F_s$ is the sticktion force, and $v_s$ is the Stribeck velocity.

The simulation model of the LuGre model, as shown in Figure 6, is built in Simulink with one input signal that is a relative velocity from sprung mass and one output signal that is control force $F_{mr}$ for the suspension system connected with a summation junction of the sprung and un-sprung masses with (-) and (+) signs respectively alike Bingham and Dahl models shown in Figure 4 and 5.

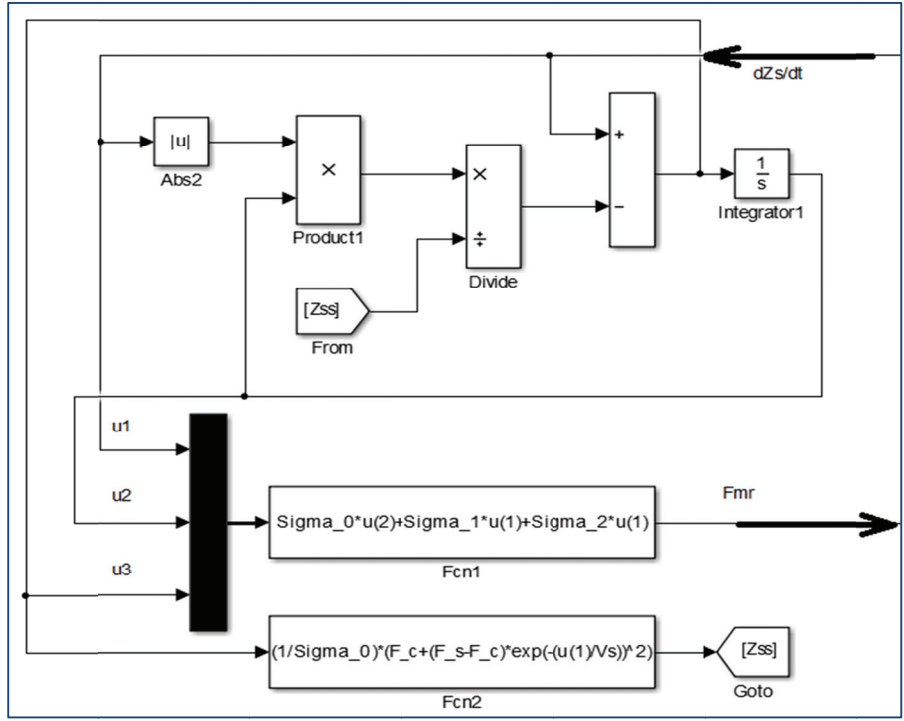

Figure 6. Simulink model of the LuGre model.

In the Simulink model, a function block with three input signals, viz. $\dot{z}(t), y(t), \dot{y}(t)$, is employed to compute a control force that is the MR force $F_{mr}$. The two input signals, which are $y(t)$ and $\dot{y}(t)$, are internal variables computed from the expressions (7) and (8).

### 3.4. Bouc-Wen model

The MR damper with Bouc-Wen model is composed of stiffness (spring) element, passive damper and Bouc-Wen hysteresis loop elements. The schematic representation of Bouc-Wen model of an MR damper is depicted by the next schematic view – Figure 7. The hysteresis loop has an internal variable $y$ that represents hysteretic behavior and satisfies the next expression (9). The model equation of Bouc-Wen model [8] is expressed by the following.

$$\dot{y} = -\gamma |\dot{z}| \, y \, |y|^{n-1} - \beta z \, |y|^n + A\dot{z} \qquad (9)$$

Where $y$ is the evolutionary variable that can vary from a sinusoidal to a quasi-rectangular function of the time depending on the parameters $\gamma, \beta$ and $A$.

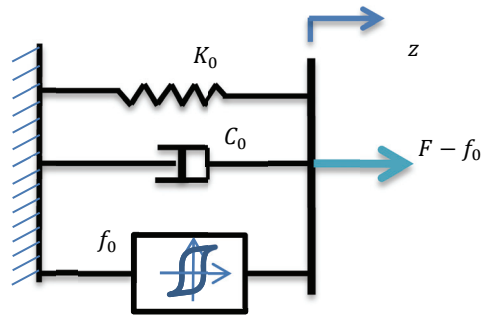

Figure 7. Schematic representation of Bouc-Wen model of an MR damper.

The force exerted by the MR damper is the function of the relative displacement $z$ and velocity $\dot{z}$ and the parameter $\alpha$ defined by the control voltage $u$, and is given by

$$F_{mr} = C_0(u)\dot{z} + K_0 z + \alpha(u)y + f_0 \qquad (10)$$

In the model computing damping force of the MR damper, $K_0$ is the stiffness of the spring element of the MR damper and the values of the parameters (coefficients) $C_0(u)$ and $\alpha(u)$ have a linearly relationship with the control voltage $u$ and determine the influence of the model on the final force $F$. The force $f_0$ takes into account pre-yield stress of the damper. The values of the parameters (coefficients) $C_0(u)$ and $\alpha(u)$ are determined from the following expressions:

$$C_0(u) = C_{0a} + C_{0b}u, \quad \alpha(u) = \alpha_{0a} + \alpha_{0b}u \qquad (11)$$

The best fit parameter values of these parameters are determined by fitting to the experimentally measured response of the system.

The simulation model of the system from Bouc-Wen model shown in Figure 8 is built in Simulink by using the equations expressed in (9), (10) and (11). The simulation model has two input sources, viz. $z(t)$ displacement and $dz(t)$ velocity of the sprung mass ($m_s$) of the system, and two output signals for control force $F_{mr}$ going to the sprung mass ($m_s$) with (-) minus sign and to the un-sprung ($m_u$) mass with (+) plus sign. Note that $dz(t)$ is equal to $\dot{z}(t)$ and $F_{mr}$ is equal to $U_c$ in the equation (2). Note that in the MR model, there are two input signals and one output signal. The input signals are $z(t)$ and $\dot{z}(t)$ displacement and velocity of the sprung mass and the output signal is the control force $F_{mr}$ generated by the MR damper.

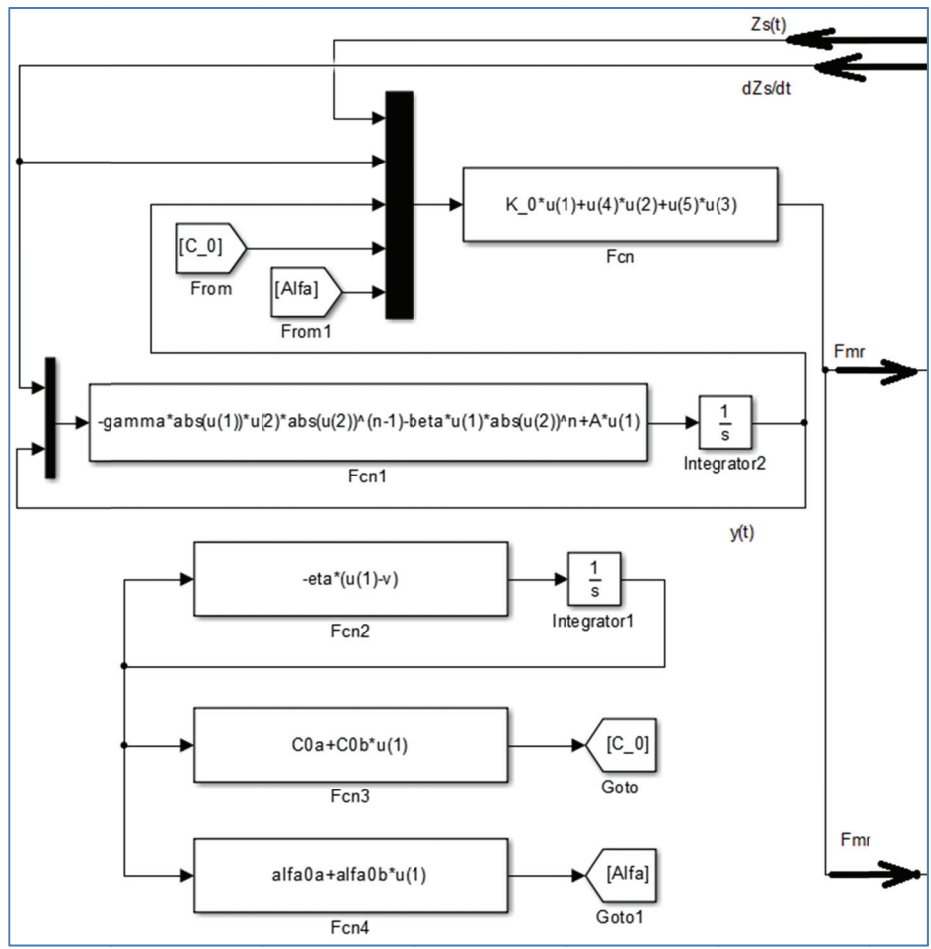

Figure 8. Bouc-Wen model.

In Bouc-Wen model alike Bingham, Dahl and LuGre models, the control force feeds the summing junction of forces for the sprung mass with (-) sign and for the un-sprung mass with (+) sign.

Also, all of the four simulation models are summed up as sub-systems (Figure 9) to compare their performances against each other and a passively controlled system for four different excitation signals from the terrain. The system response is displacement in the car body from the road excitations.

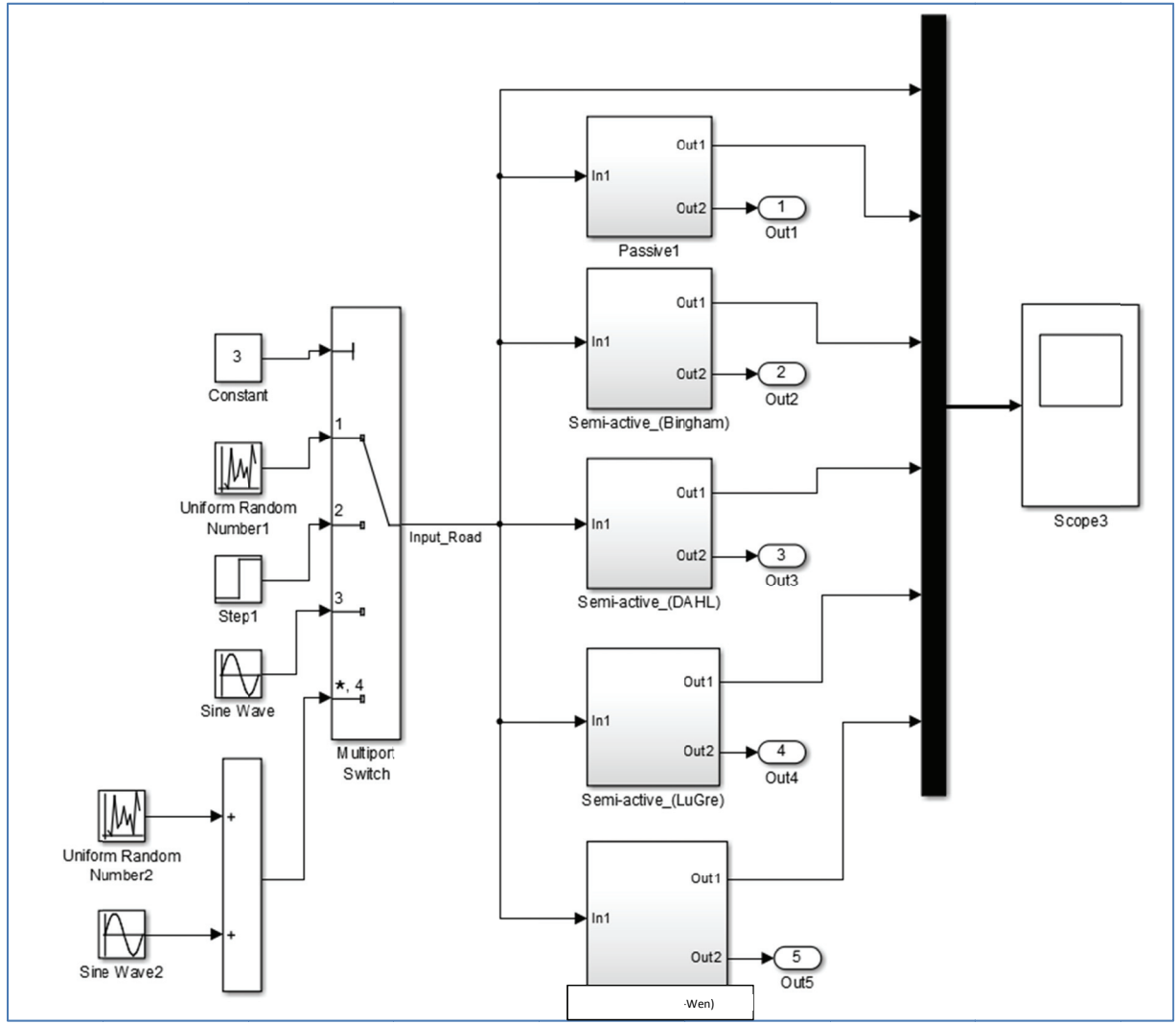

Figure 9. Passively controlled system model vs. four MR models as sub-systems.

4. **Simulation results and discussions**

The above depicted mathematical formulations as implemented in Simulink models are simulated to compare performances of each model with respect to its exerted damping force, and vibration and shock damping efficiency as a semi-active vibration controller formulated in the system equations (2) of motion against passively controlled/damped vibration damper formulated in (1) in the example of quarter car model shown in Figure 1. In all of our simulations, the control force in (2) $U_c$ is set to be equal to $F_{mr}$ and vibration damping is evaluated in the sprung mass. Displacement values of the sprung mass with a semi-active controller of the MR damper models are compared with the displacement values of a passively controlled suspension system. The values of suspension parameters (quarter car) are taken from the data given in Table 1 and all numerical values for hysteresis model (Bingham, Dahl, LuGre and Bouc-Wen) parameters are chosen from the data given in Table 2, 3, 4 and 5. The rational parameter values of the hysteresis models are found by trails and errors. For numerical simulations three different signals, viz.

random white noise, Heaviside step function and sine waves with 2.1 Hz and 20.8 Hz of oscillations, and also, a combinatorial excitation signal, a sum of sine waves and random (Gaussian white) noises, are taken. Road excitation signals are set to have maximum (absolute) magnitude of 0.075 m and oscillation frequencies of sine waves are taken by considering natural frequencies of the quarter car model.

**Table 1.** Data for suspension system (quarter car model).

| Parameter name | Parameter notation | Parameter value |
|---|---|---|
| Sprung Mass | $m_s$ | $2500\ kg$ |
| Un-sprung mass | $m_u$ | $320\ kg$ |
| Stiffness of suspension | $k_s$ | $80000\ [N/m]$ |
| Stiffness of un-spring mass (tire) | $k_u$ | $500000\ [N/m]$ |
| Damping coefficient of sprung mass | $c_s$ | $320\ [N \cdot s/m]$ |
| Damping coefficient of un-sprung mass | $c_u$ | $15020\ [N \cdot s/m]$ |

**Table 2.** Data for Bingham model simulation.

| Parameter name | Parameter notation | Parameter value |
|---|---|---|
| Damping coefficient in Bingham model | $c_0$ | $320\ [N \cdot s/m]$ |
| Offset force | $F_0$ | $10\ N$ |
| Frictional force | $F_c$ | $100\ N$ |

**Table 3.** Data for Dahl model simulation.

| Parameter name | Parameter notation | Parameter value |
|---|---|---|
| Control voltage | $v$ | $5\ [V]$ |
| Hysteresis parameters | $k, k_{wa}, k_{wb}, \rho$ | 350, 800, 250, 25 |

**Table 4.** Data for the LuGre model simulation.

| Parameter name | Parameter notation | Parameter value |
|---|---|---|
| Coulomb friction force | $F_c$ | $10\ [N]$ |
| Sticktion force | $F_s$ | $25\ [N]$ |
| Stribeck velocity | $v_s$ | $0.04\ [m/s]$ |
| Stiffness coefficient | $\sigma_0$ | $500\ [N/m]$ |
| Damping coefficient | $\sigma_1$ | $10^4\ [N \cdot s/m]$ |
| Viscous friction coefficient | $\sigma_2$ | $0.6\ [N \cdot s/m]$ |

**Table 5.** Data for Bouc-Wen model simulation.

| Parameter name | Parameter notation | Parameter value |
|---|---|---|
| Parameters of the Hysteresis shape | $\gamma, \beta, A, n$ | 1, 0, 1.5, 2 |
| Stiffness of the spring element | $K_0$ | $300\ [N/m]$ |
| Input voltage $u$ | $v$ | $5\ [V]$ |
| Other parameters | $C_{0a}, C_{0b}, \alpha_{0a}, \alpha_{0b}$ | 4400, 442, 10872, 49616 |
| Pre-yield stress | $f_0$ | $0\ [N]$ |

From the numerical simulations of hysteresis loop models with Bingham, Dahl, LuGre and Bouc-Wen models for the semi-active suspension system it is clear that all of the semi-active system models outperform passively controlled system model for four different excitation signals from road. Figure 10 and 11 demonstrate system responses (displacement of the car body) of the passively and semi-actively controlled models from random (Gaussian white) noise with the magnitude of 0.075 m (in the range of -0.0375 m … +0.0375 m) and from the simulation results it is clearly seen that all hysteresis models outperform passively damped system model in damping undersigned excitations from the terrain. Out of these four semi-active models, Bingham and Bouc-Wen models demonstrate much higher damping than the other two models, viz. Dahl and LuGre models.

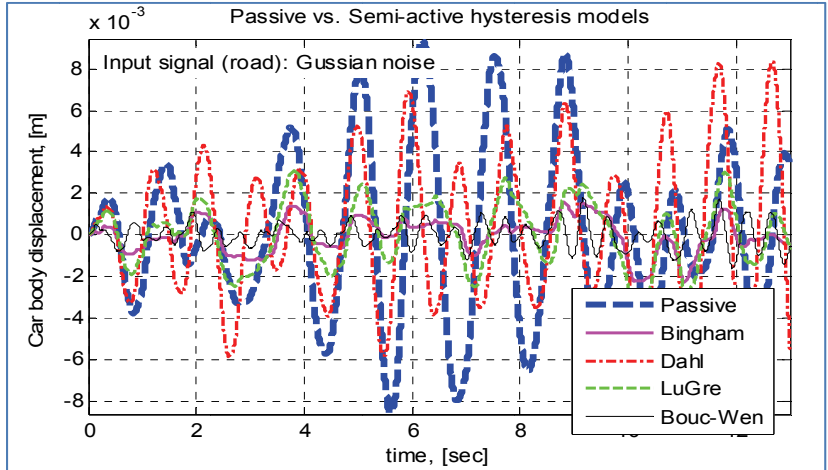

Figure 10. Model responses on random (Gaussian white noise) excitation from road.

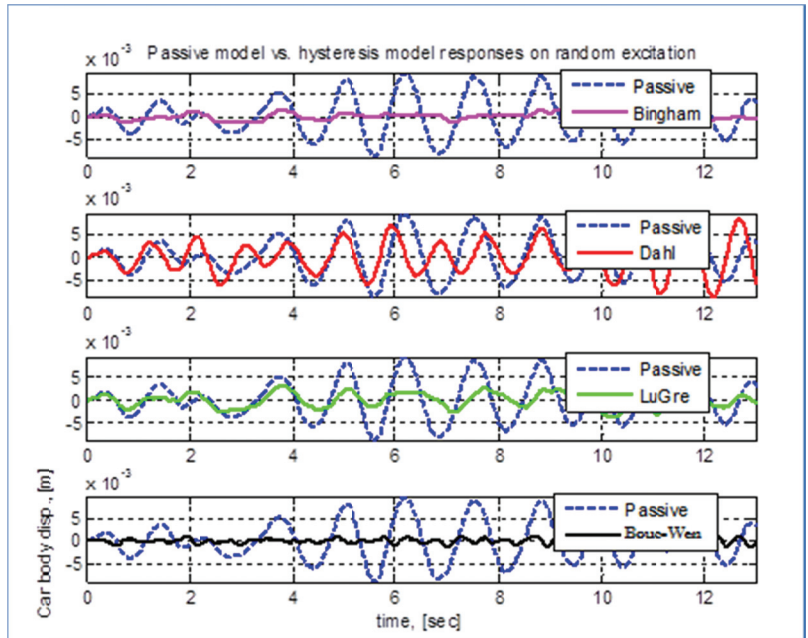

Figure 11. Model responses on random (Gaussian white noise) excitation from road.

In another excitation with Step (Heaviside) function – Figure 12 and 13, the hysteresis models outperform in damping undesired excitation in the car body in comparison with passively controlled model. In this case, LuGre and Bouc-Wen models performs considerably better than the other two models and dissipate the step excitation with the magnitude of 0.075 m in less than 2 seconds. Whereas Bingham model damps the excitation in about 8 seconds and Dahl model in about 13 seconds.

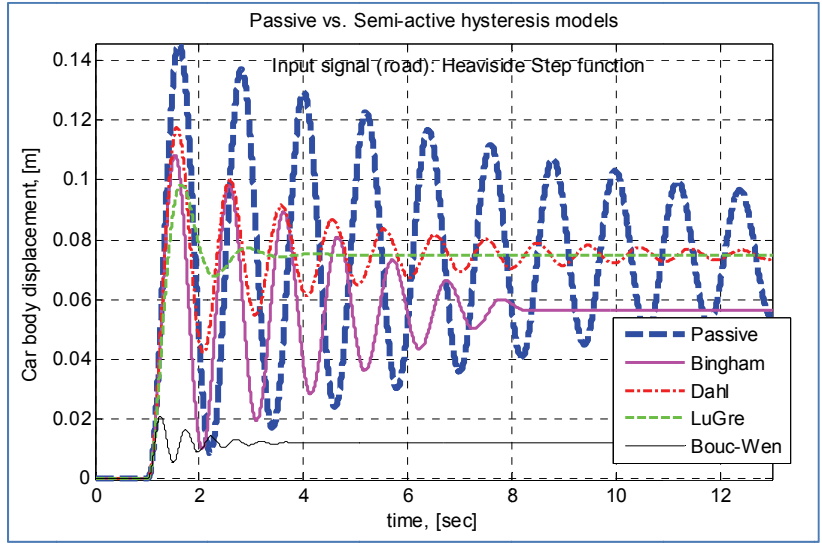

Figure 12. Passive vs. semi-active suspension models on Step input excitation.

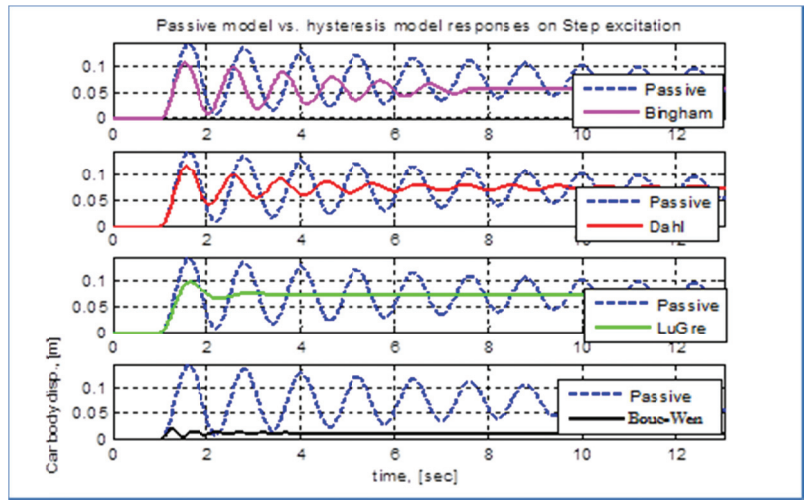

Figure 13. Passive vs. semi-active suspension models on Step input excitation.

In sine wave excitations with 2.1 Hz of frequency shown in Figure 14 and 15, semi-actively controlled models except for Bouc-Wen model have demonstrated slightly better in damping magnitude of excitation oscillations in comparison with a passively controlled system model and frequency of excitation from the road is preserved clearly as a periodic signal with all models. In this case, Bouc-Wen model has outperformed all other models in terms of damped oscillation magnitudes. In sine wave excitations with 20.8 Hz of frequency shown in Figure 16 and 17, all hysteresis models have dissipated magnitude of excited vibrations in car body more than passively controlled model by preserving periodic oscillations with respect to road excitations. Performances of all models after about two seconds of simulation time have reached to very similar steady state value in the range of $\pm 5\ mm$ of displacement in car body. In this case, Bouc-Wen model has performed slightly poorer than the other three MR damper models in terms of damped excitation.

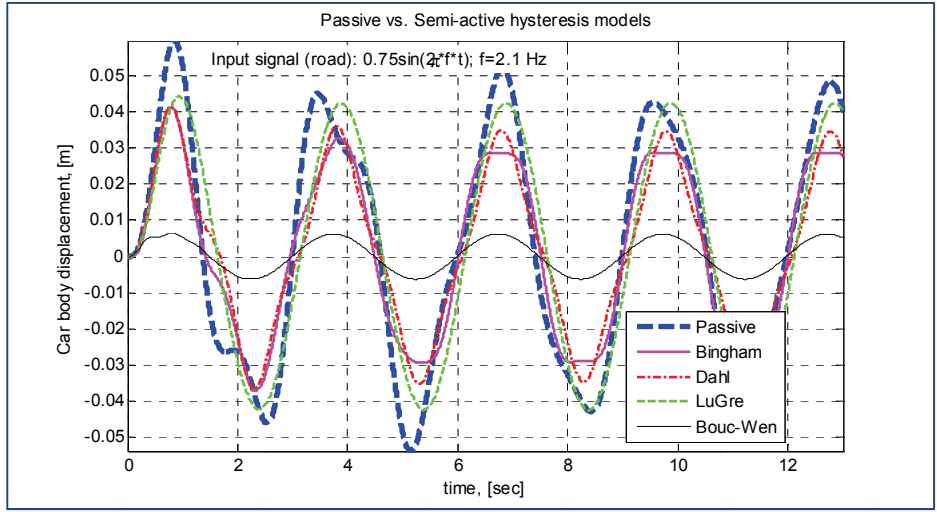

Figure 14. Passive vs. semi-active suspension on sinusoidal wave: $0.075\sin(2\pi ft), f = 2.1\ Hz$ excitation.

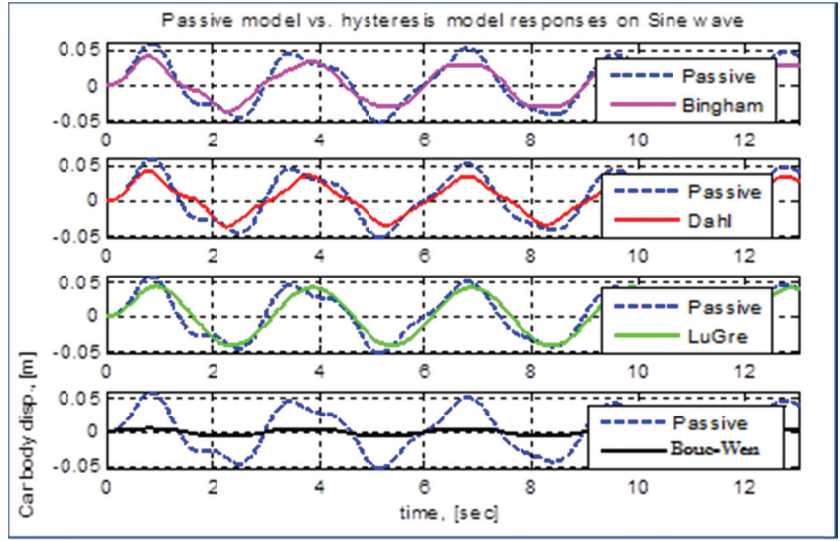

Figure 15. Passive vs. semi-active suspension on sinusoidal wave: $0.075\sin(2\pi ft), f = 2.1\ Hz$ excitation.

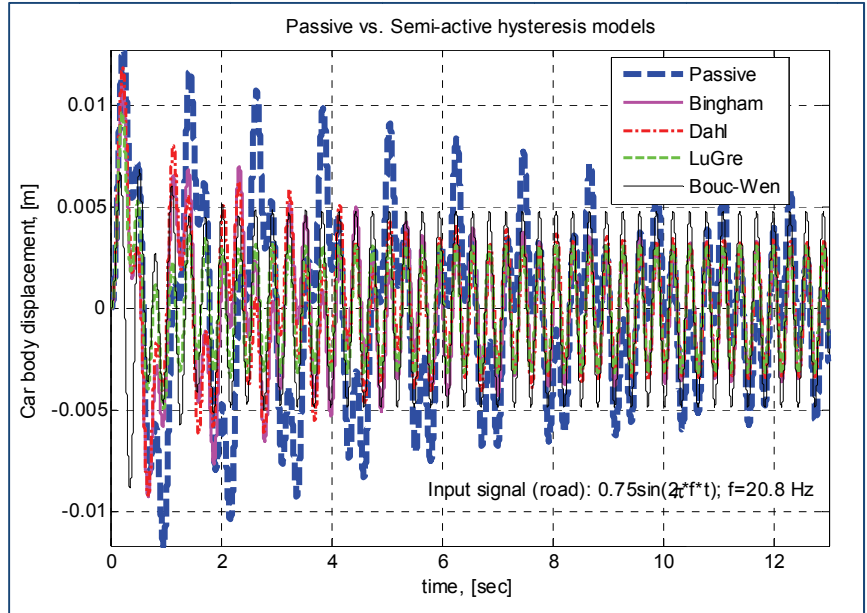

Figure 16. Passive vs. semi-active suspension on sinusoidal wave: $0.075 \sin(2\pi f t), f = 20.8\ Hz$ excitation

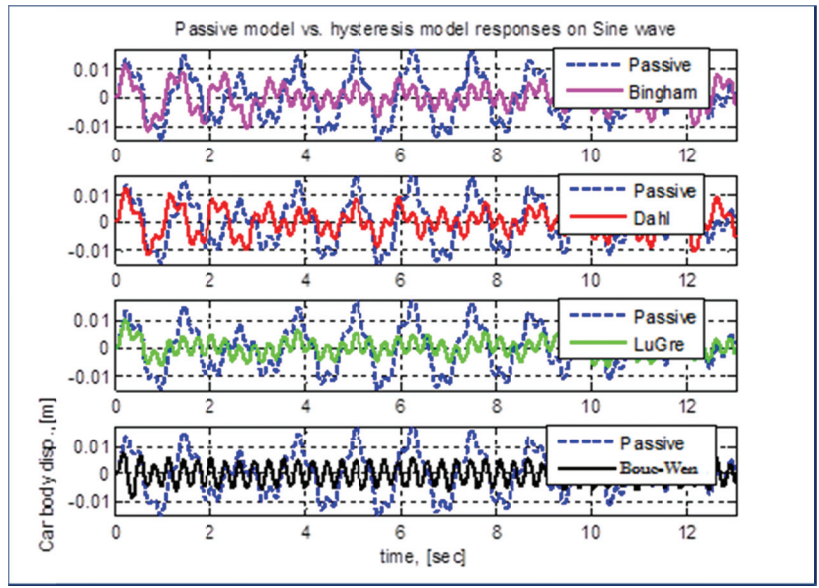

Figure 17. Passive vs. semi-active suspension on sinusoidal wave: $0.075 \sin(2\pi f t), f = 20.8\ Hz$ excitation

A fourth excitation signal from road used to simulate the models is sine wave with 20.8 Hz of frequency plus white noise. The performances of the semi-active models for this excitation – Figure 18 and 19 have been similar to the previous case with sine wave excitation with 20.8 Hz frequency for some extent and in this case, the two MR models, viz. Bingham and Dahl models, have not reached to a stable steady-state value.

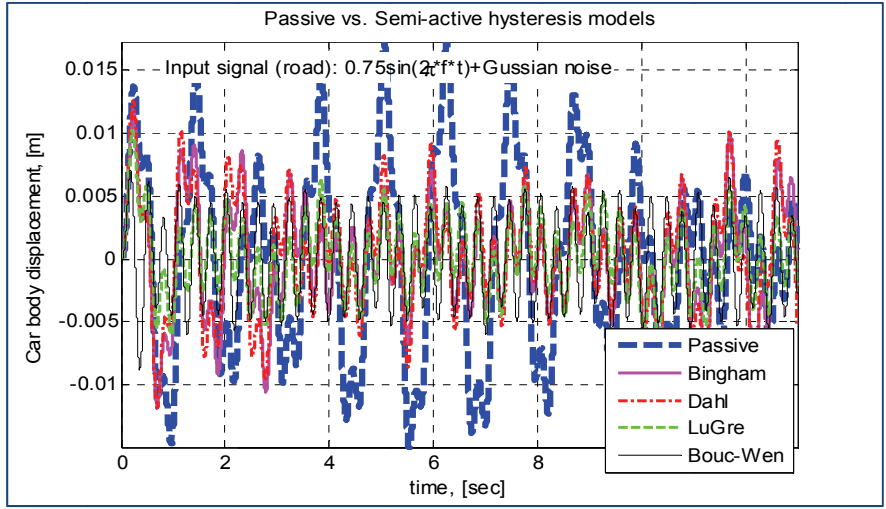

Figure 18. The system responses of passive and semi-active on sinusoidal wave ($f = 20.8 Hz$): $0.075 \sin(2\pi f t)$ + Gaussian white noise excitation.

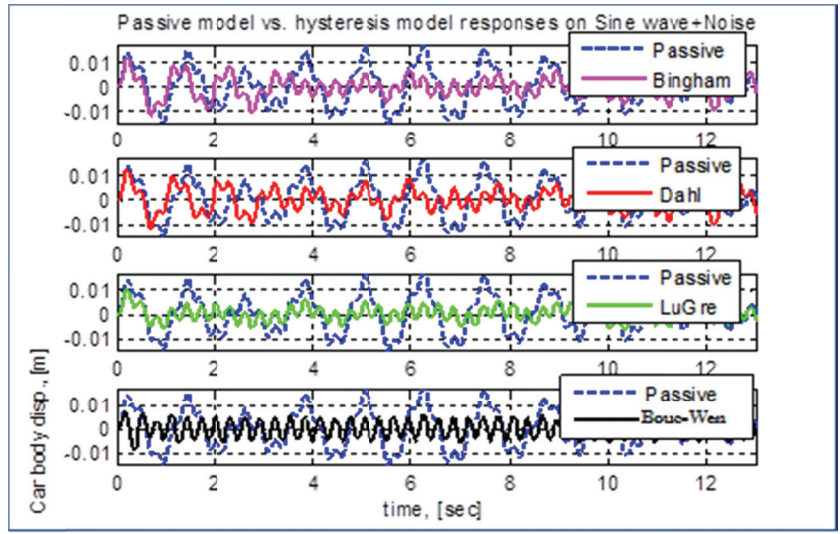

Figure 19. The system responses of passive and semi-active on sinusoidal wave ($f = 20.8 Hz$): $0.075 \sin(2\pi f t)$ + Gaussian white noise excitation.

## 5. Summary

The developed simulation models of the hysteresis or non-linear system behaviors of the MR liquids used in dampers by using mathematical formulations of Bingham, Dahl, LuGre and Bouc-Wen models in MATLAB/Simulink in the example of quarter car model have showed adequacy of these MR dampers for designing vibration and shock dampers. The simulation results of the semi-actively controlled damper models with Bingham, Dahl, LuGre and Bouc-Wen models have demonstrated superiority over passively controlled damper model in the example of four different excitation signals mimicking terrain roughness for the quarter car suspension system model. Amongst these semi-actively controlled models, Bouc-Wen has outperformed other models in terms of the damped vibrations and steady-state response time in three excitation signals, viz. step, white noise and low frequency sine wave. In case of higher (>20.8 Hz) frequency (pure periodic) excitation from road, Bingham, Dahl and LuGre models perform better than Bouc-Wen model.

Further studies will be aimed to develop mathematical (empirical) formulations and experimental validations to compute optimal parameters of MR hysteresis based dampers with Dahl, LuGre and Bouc-Wen models with respect to suspension and tire parameters. In addition, it is planned to develop an adaptive PID controller in association with these MR damper models.

**Acknowledgements.** This research study is supported by the state grant # A-3-54 from the State Science and Technology Committee of Uzbekistan.

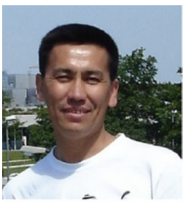
**Sulaymon L. ESHKABILOV.** He got ME from Tashkent Automotive Road Institute (TARI) in 1994, earned MSc from Rochester Institute of Technology in 2001 and PhD from Cybernetics research Institute of Academy Sciences of Uzbekistan in 2005. He was a visiting professor in 2010-2011 at Mechanical Engineering department of Ohio University, OH, USA. He is currently an associate professor at TARI and holds a part-time professor position at Tashkent-Turin Polytechnic University. His research areas are mechanical vibrations and applications of modal analysis, system identification, control and modeling of dynamic systems.